\documentstyle [12pt] {article}

\begin{document}
\begin{flushright}
IUHET-249\\
IUNTC93-14\\
May 1993\\
\end{flushright}
\vspace*{0.5 cm}
\begin{center}
{\bf LONG RANGE FORCES\\
FROM\\
THE COSMOLOGICAL NEUTRINO BACKGROUND}\\
\vspace*{0.5 cm}
\vspace*{0.5 cm}
{\bf C. J. Horowitz and J. Pantaleone}\\
\vspace*{0.5 cm}
Physics Department\\
Indiana University, Bloomington, IN 47405 \\
\vspace*{0.5 cm}
{\bf ABSTRACT} \\
\end{center}

The cosmological neutrino background will mediate long range forces
between objects.
For a background of temperature T,
the potential decreases as \(1/r^5\) for \(r >> 1/T\)
and as \(1/r\) for \(r << 1/T\).
These forces have large spin-dependent components.
If the neutrino background is nonrelativistic,
the long range forces are enhanced by a
factor of the inverse neutrino velocity.
These long range forces may provide a method for
observing the cosmological neutrino background.

\newpage
Ever since the neutrino was suggested by Pauli,
there have been speculations about the long range forces produced
by neutrino exchange (\cite{FS,HS} and references therein).
Feinberg and Sucher found a  potential which falls off like
\(1/r^5\) with distance from two neutrino exchange in vacuum.
Exchange of an even number of neutrinos is required in vacuum
in order to leave the spin of the external particles unchanged,
so that the effects of the potential can sum coherently over many particles.
However in a background of neutrinos, single neutrino exchange
can mediate a coherent, long range force
because the background can supply and absorb one of the neutrinos.
This is a realistic possibility since a large,
cosmological neutrino background is firmly believed to exist
by analogy with the observed cosmological photon background.
The neutrino background has not been directly observed,
but is the subject of considerable interest since it
is a favorite candidate for dark matter.
We shall calculate the long range forces from neutrino exchange
in the cosmological neutrino background.

\section{Calculation of Potentials.}

We will use the formalism of finite temperature field theory
\cite{FTFT,FW,HW,NR} to calculate the Feynman diagram shown in
Fig. (1).  We assume that there are two fermions
which interact with neutrinos via the effective Langrangian
\begin{equation}
L_{int} = - \sqrt{2} G_F {\bar \nu} \gamma^\mu {1 \over 2} (1 - \gamma^5) \nu
\ \ {\bar F} \gamma_\mu (g_V - g_A \gamma^5) F
\label{coupling}
\end{equation}
where \(F\) represents either of the two fermions.
\(g_V\) and \(g_A\) are
composition dependent charges, of order 1,
which can depend on the flavor of the neutrino and, for massive
neutrinos, on the lepton mixing parameters (see Table).

We  assume that the two fermions are both nonrelativistic in
the rest frame of the medium.
Then the scattering amplitude, \(T(Q)\), of the two fermions is
\begin{equation}
iT(Q) = (-i \sqrt{2}G_F)^2 (-1) (g_V, -2g_A {\vec S})^\mu
 (g'_V, -2g'_A {\vec S'})^\nu \ \ I_{\mu \nu}(Q)
\end{equation}
Here \(Q \approx (0,{\vec Q})\) is the momentum transfer,
\({\vec S}\) denotes the fermion spin normalized to
\({\vec S}^2 = 3/4\),
and the factor of (-1) is for the fermion loop.
Terms down by factors of the external particles'
velocity through the medium are neglected
(however for situations when the test object's spin averages
to zero, then in the above expression
the spin vector can be replaced by the object's velocity
vector with respect to the medium:
\(-2g_A {\vec S} \rightarrow g_V {\vec v}\) ).
The potential in real space is just the fourier transform of
\(-T(\vert {\vec Q} \vert )\),
\begin{equation}
V(r) = \int {d^3 \vec{Q} \over (2\pi)^3} \ \
e^{i \vec{Q} \cdot \vec{r}} \ \
( -T(\vert {\vec Q} \vert ) )
\label{FT}
\end{equation}
The \(I^{\mu \nu}\) is given by
\begin{equation}
I^{\mu \nu} = \int { {d^4 k} \over { (2 \pi )^4}}
Tr[ \gamma^\mu P_L i S_T(k) \gamma^\nu P_L i S_T(k-Q) ]
\end{equation}
where \(P_L\) is the projection operator for the left-handed chirality.
The neutrino propagator is
\begin{equation}
S_T(k) = ( k_\mu \gamma^\mu + m )
[ { 1 \over {k^2 - m^2 + i \epsilon}} + i 2 \pi \delta (k^2 -m^2)
[\Theta(k^0)n_+ + \Theta(-k^0)n_-] ]
\label{S_T}
\end{equation}
\(n_{\pm}\) are the +, neutrino and -, antineutrino
distribution functions \cite{dense}.
\(m\) is the neutrino's Dirac mass, the possibility of Majorana
neutrino masses is not considered here.

This expression for \(I^{\mu \nu}\) correctly gives
the {\it real} part of the amplitude,
but not the {\it imaginary} part.
In general, a proper treatment of the
imaginary part of the vacuum polarization requires different
Feynman rules (see e.g. \cite{FW,HW}).
However we are only interested in the static limit
where \(Q^0 = 0\) and then the imaginary part of the
retarded amplitude must vanish.  This is because in the
the static limit the background can not emit or absorb energy.

The first part of Eq. (\ref{S_T}) is the vacuum propagator.
Using only this part from each neutrino propagator
gives for the spin-independent long range potential,
in vacuum, in the large r limit,
for m=0,
\begin{equation}
V_{SI,vac}(r) = { {G_F^2 g_V g'_V} \over {8 \pi^3}} { 1 \over {r^5}}
\label{V_vac}
\end{equation}
This agrees with the recent recalculation by Hsu and Sikivie \cite{HS}.

The thermal contribution to the long range force
comes from the "cross terms" when one neutrino propagator is the
vacuum part and the other neutrino propagator is the thermal part.
This corresponds to a background neutrino being excited to
a virtual state by one scattering and then, with the other scattering,
being returned to the background with its original energy and momentum.
This part of \(I^{\mu \nu}\) can be written as
\begin{eqnarray}
I_T^{\mu \nu} & = & \int { {d^4 k} \over { (2 \pi )^4}} (-i) 2 \pi
\ \ \delta(k^2 - m^2)
[\Theta(k^0)n_+ + \Theta(-k^0)n_-] ] \nonumber \\
& & \mbox{} \{
{Tr[ \gamma^\mu P_L  ({\not k}+{\not Q}) \gamma^\nu P_L {\not k} ] \over
(k+Q)^2 - m^2 + i\epsilon} +
{Tr[ \gamma^\mu P_L  {\not k} \gamma^\nu P_L ({\not k}-{\not Q}) ] \over
(k-Q)^2 - m^2 + i\epsilon}
\}
\label{I2}
\end{eqnarray}
This expression will be evaluated in certain limits.
Typically, the above integrations can be easily done in the following order:
first the integration over \(k^0\), then the fourier transform, Eq. (\ref{FT}),
then the angular integration in neutrino momentum space.  This leaves a final
integration over the background neutrino momentum which depends on their
distribution functions.  For simple distribution functions this integration
can be done analytically to yield compact results.

\subsection{Cosmological Neutrinos.}

For cosmological neutrinos
we use the Boltzmann approximation
for the background neutrino distribution functions.
Then the integration over neutrino momentum can be done analytically.
Assuming only one species of background neutrino,
\begin{equation}
n_{\pm} = { \rm exp}[ (\pm \mu - \vert {\vec k}\vert ) / T ]
\end{equation}
where \(T\) is the neutrino temperature,
\( T = (4/11)^{(1/3)} T_\gamma \approx 1/1.2\)mm and
\(T_\gamma\) is the photon temperature.
The neutrino chemical potential, \(\mu\), is expected
to be small compared to the temperature, \(T\).
We generally take the chemical potential to be zero,
except when calculating one of the parity violating potentials.

If the background neutrino is massive, with a mass less than a few MeV's,
then this distribution function is still a good approximation \cite{LLS}!
This is because then the neutrinos decoupled when they were relativistic,
and the expansion of the universe preserves the momentum distribution.
However if the massive neutrinos form clumps,
then this distribution will be modified.
Then the local (galactic or galactic cluster) neutrino
energy density could be enhanced by several orders of magnitude.

For m=0, the spin-independent potential from excitation of the
neutrino background is
\begin{equation}
V_{SI,0}(r) = { { - 8 G_F^2 g_V g'_V} \over {\pi^3}} { 1 \over r}
{T^4 \over ( 1 + (2rT)^2 )^2}
\label{V_0}
\end{equation}
For massive neutrinos, vacuum potentials are exponentially suppressed
at large distances reflecting the small probability for
a massive particle-antiparticle pair to tunnel out of the vacuum.
However in an existing background of massive neutrinos, there
is no exponential suppression
since the neutrinos mass energy is  already present.
For \(m >> T\), the spin-independent potential is
\begin{equation}
V_{SI,m}(r) = { { - 2 G_F^2 g_V g'_V} \over {\pi^3}} { 1 \over r}
{m T^3 \over ( 1 + (2rT)^2 )^2}
\label{V_m}
\end{equation}
where we have only kept the term enhanced by the large
numerical factor of m/T,
the inverse of the background neutrino's velocity.

The above potentials, Eqs. (\ref{V_0}) and (\ref{V_m}),
fall off like  1/r\(^5\) for r \(>>\) 1/T, and go like 1/r for r \(<<\) 1/T.
A heuristic explanation of the latter behavior is possible.
Over short distances all of the background neutrinos
will be excited to a virtual state and hence
can similarly contribute to the long range interaction.
In particular,
then all the propagator dynamics factor out of the integration over
neutrino momentum in Eq. (\ref{I2}).
The potentials at \(r << 1/T\) are proportional to the background
neutrino energy density (also Eq. (\ref{DENSE})).
Then, dimensionally, the force must go like \(1/r\).

There are many neutrino background induced, spin-dependent, potentials.
The massive neutrino case is especially interesting
since at large distances it is enhanced
by the factor of m/T.  The spin-dependent piece for \(m >> T\) is
\begin{equation}
V_{SD,m}(r) = (2 g_A {\vec S}) \cdot (2 g'_A {\vec S'})
{V_{SI,m}(r) \over g_V g'_V }
\end{equation}
When one of the spins averages to zero, the velocity of that particle
enters the above expression and the potential is then parity violating
(see discussion below Eq. (\ref{coupling})).
Parity violating potentials do not occur in gravity or electro-magnetism.
There is also a velocity independent parity violating potential in a
neutrino background.  For the massive or massless neutrino case
\begin{equation}
V_{PV}(r) =  {\hat r} \cdot ( {\vec S} \times {\vec S}')
{ {  8 G_F^2 g_A g'_A} \over {\pi^3}} T^3
\sinh({\mu \over T})
{ ( 1 + 5 (2rT)^2 ) \over r^2 ( 1 + (2rT)^2 )^3}
\end{equation}
Note that this falls off more rapidly with r
than the previous potentials.
Also, it vanishes if the chemical potential vanishes,
and so is expected to be suppressed.
The observed baryon asymmetry is
\( \eta = n_b-n_{\bar b}/n_\gamma \approx 10^{-9}\),
so one naively expects a similar asymmetry for the neutrinos.
However \(\sinh(\mu / T)\) is presently unconstrained by observations,
it could easily be as large as \(10^{-2}\)
or larger \cite{Olive}.

\subsection{Supernova Neutrinos.}

In a supernova, neutrinos can be degenerate, \(\mu/T >> 1\).
Then the antineutrino background vanishes
and the distribution function for the neutrino background
is just
\[
n_+  \approx \left\{
\begin{array}{ll}
1  & \mbox{} 0 < k^0 < \mu \\
0  & \mbox{} \mu < k^0
\end{array} \right. \]
Then the spin independent potential is, for m=0
\begin{equation}
V_{SI,\mu}(r) = { { -  G_F^2 g_V g'_V} \over {4 \pi^3}} { 1 \over r^4}
[ {1 \over r} - {\cos(2 \mu r) \over r} - \mu \sin(2 \mu r) ]
\label{DENSE}
\end{equation}
At long distances, \(r >> 1/\mu\), the potential oscillates in
sign with a \(1/r^4\) fall off, while for \(r << 1/\mu\)
it goes like 1/r.

The oscillations in Eq. (\ref{DENSE}) are related to the
(assumedly) sharp Fermi surface.  They are directly
analogous to Friedel oscillations \cite{FW,Friedel}
of the electron screening charge in a metal.
Presumably as the temperature is increased, and the sharp
Fermi surface smears out, the oscillations will become
exponentially damped.

\section{Applications.}

\subsection{Detecting Cosmological Neutrinos.}

There has been considerable experimental interest lately in
searching for small, exotic forces (see e.g. \cite{Adel}).
The recent observation of
the Casimir-Polder potential \cite{CP,meas}, which falls off like
\(1/r^7\),  offers some mild encouragement for the possibility
of future observations of neutrino exchange potentials.
While the spin-independent forces between atoms
from neutrino exchange
are typically much smaller than those from
electromagnetism and gravity, the spin-dependent
and parity violating forces are not.
Thus the neutrino background induced spin-dependent potentials
may prove to be experimentally accessible.
They could provide a good technique for detecting the
cosmological neutrino background.

Many effects have been previously discussed for observing
cosmological neutrinos.
One often discussed detection technique is to look for forces
from coherent scattering at surfaces \cite{Opher}.
These forces are naively first order in \(G_F\),
however they are suppressed down to \(G_F^2\) size for
objects in a uniform neutrino background \cite{LLS}.
Also, they vanish in a neutrino-antineutrino symmetric background.
Thus these surface effects do not look promising (see e.g. \cite{Smith}).

A better detection technique is to look for the drag deceleration
due to the cosmological neutrino background.
For macroscopic objects smaller than 1/T,
the neutrinos scatter coherently from the objects volume.
Then if the object is moving through the neutrino background,
it will experience a composition dependent drag.
This effect may be observable \cite{russian}.

The long range forces induced by the cosmological neutrino background
act between two objects, and so
are qualitatively different then the previously discussed forces
\cite{Opher,russian}.
However it is encouraging to note that they do also
sum coherently over 1/T size target volumes
and do not vanish in a neutrino-antineutrino symmetric background.
In addition, they may have several experimental advantages
over previous proposals.
The dependence on r provides an additional
experimental handle which can be used to vary the size of the effect.
The effect is present for an experiment at rest in the neutrino background.
Also, at short distances the induced forces are directly proportional to the
background neutrino energy density--an extremely
likely candidate for dark matter.
Gravitational clumping of a massive neutrino background
could increase the local neutrino energy density
by several orders of magnitude.  This would enhance these forces.

\subsection{Astrophysics.}

In the cosmological neutrino background, the distance between
neutrinos is roughly \(r \approx 1/T\).
Comparing the gravitational and background induced
potentials (see also \cite{DF}) for neighboring,
massive background neutrinos,
\begin{equation}
{V_{Gravity} \over V_{SI,m}} \approx
({ m \over 30 eV}) ( {10^{-2} eV \over T})^3
\end{equation}
Thus the interaction between neighboring neutrinos in the cosmological
background is dominated by the neutrino exchange
contribution, up until almost present temperatures.
However this interaction is always smaller than the temperature,
so large effects in the dynamics do not appear likely.

\section{Conclusions.}

We have calculated the long range forces induced by a neutrino background.
These potentials go like \(1/r\) over distances shorter than \(1/T\),
but fall off more rapidly beyond that.
They can be enhanced by a factor of \(m/T\) (about \(10^5\)
for a 30 eV neutrino mass) over the forces
from massless neutrino exchange in vacuum.
But still the spin-independent forces between atoms
are typically very small compared
to the gravitational and electromagnetic spin-independent forces.
However the neutrino background induced spin-dependent (and parity violating)
forces are relatively large and may provide a method for observing
the cosmological neutrino background.
This method may have several advantages over previously proposed methods,
but it still appears to be an extremely challenging experiment.

\begin{center}
Acknowledgements
\end{center}

We would like to thank A. Hendry, A. Kostelecky, D. Lichtenberg
and S. Samuel for useful discussions.
This work is supported in part by
the U.S. Department of Energy under Grant No. DE-FG02-91ER40661.

\raggedbottom

\raggedbottom
\newpage

Table 1.  Values for \(g_V\) and \(g_A\).
Here \(G_A \approx 1.25\) is the axial vector form factor,
and \(x = \sin^2 \theta_W \approx 0.23\).
When the external particle is the antiparticle,
\(g_V\) changes sign while \(g_A\) does not.
\\

\begin{tabular}{| c | c | c | c |} \hline
External Flavor & Internal Flavor & \(g_V\) & \(g_A\) \\ \hline
{\rm e} & \(\nu_e\) & \({1 \over 2} + 2 x \) & \({1 \over 2}\) \\
{\rm e} & \(\nu_\mu , \nu_\tau\) &
\(- {1 \over 2} + 2 x \) & \(- {1 \over 2}\) \\
{\rm p} & \(\nu_e , \nu_\mu , \nu_\tau\) &
\({1 \over 2} - 2 x\)  & \({G_A \over 2}\) \\
{\rm n} & \(\nu_e , \nu_\mu , \nu_\tau\) &
\(- {1 \over 2}\)  &\( - {G_A \over 2}\) \\
\(\nu_a\) & \(\nu_a\) & 1 &  1  \\
\(\nu_a\) & \(\nu_b , a {\not =} b\) & \({1 \over 2}\)
& \( { 1 \over 2}\)  \\ \hline
\end{tabular}

\raggedbottom
\newpage

\vspace*{0.6cm}

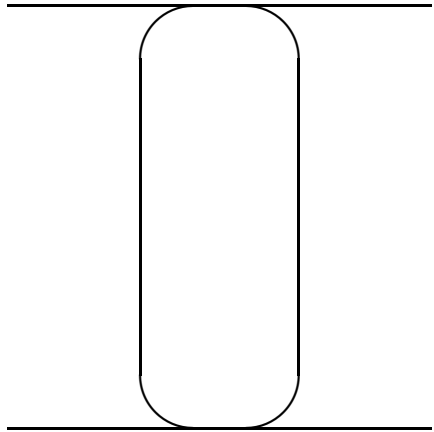
\noindent {\bf Fig. 1.} Lowest order Feynman diagram for
two neutrino exchange in the four fermion effective theory.

\vspace*{0.6cm}

\begin{picture}(200,300)

\thicklines
\put(150,100){\oval(60,160)}
\put(70,180){\line(160,0){160}}
\put(70,20){\line(160,0){160}}

\end{picture}


\begin{thebibliography}{99}
\bibitem{dense}
Here we are making the conventional
assumption that the background neutrinos are incoherent.
That is not always valid when the neutrinos are out of
equilibrium. See J. Pantaleone, Phys. Lett. B287 (1992) 128.
\end{thebibliography}
\end{document}